\documentclass[epj-spec]{svjour}
\usepackage{graphics}
\usepackage{dsfont}

\begin{document}

\title{Stochastic Resonance phenomena in qubit arrays}
\author{\'Angel Rivas \and Neil P. Oxtoby \and Susana F. Huelga\thanks{\email{s.f.huelga@herts.ac.uk}}}
\institute{Quantum Physics Group, STRI, Department of Physics,
Astronomy and Mathematics\\ University of Hertfordshire, Hatfield,
Herts AL10 9AB, UK.}
\abstract{We discuss stochastic resonance--like effects in the
context of coupled quantum spin systems. We focus here on an
information--theoretic approach and analyze the steady state quantum
correlations (entanglement) as well as the global correlations in
the system when subject to different forms of local decoherence. In
the presence of decay, it has been shown that the system displays
quantum correlations only when the noise strength is above a certain
threshold. We extend this result to the case of a Heisenberg
$XYZ$ exchange interaction and revise and clarify the mechanisms
underlying this behaviour. In the presence of pure dephasing, we
show that the system always remains separable in the steady state.
When both types of noise are present, we show that the system can
still exhibit entanglement for long times, provided that the pure
dephasing rate is not too large.
}
\maketitle

\section{Introduction}
The phenomenon of stochastic resonance (SR) \cite{benzi} epitomizes
the peculiar ways by which the interplay between coherent and
incoherent interactions may yield an optimized system's response, as
quantified by some suitable figure of merit, for some intermediate
noise strength \cite{sr98}. Initial studies on classical and
semiclassical systems soon extended to the quantum domain
\cite{sr98,andreas,qsr,milena,grifoni1n,grifoni2n}, while the concept of SR itself broadened
to account for an enhanced response in the presence of an optimal
noise rate with \cite{bu1,ours,Goychuk1} or without \cite{nancy,bu2,viola} an
underlying synchronization effect \cite{Goychuk2}. Our interest here focuses on this
latter situation with the additional ingredient that we will be
dealing with interacting quantum systems \cite{carrays}. In this
case, the system may display not only quantum coherence but also
quantum correlations (entanglement) across subsystems. The typical
scenario we will be analyzing is depicted, in its simplest form, in
figure 1. Two spin--1/2 particles ({\em qubits}) are coupled via a
Hamiltonian interaction $V_\mathrm{Q-Q}$ and individually driven whilst
subject to local forms of noise that we will model as independent
sets of harmonic oscillators. The driving is supposed to be weak, in
the sense that the external Rabi frequency $\Omega_j$ is constrained
to be such that $\Omega_j \ll \omega_0^j$ and $\Omega_j \sim J$, where
$\omega_0^j$ denotes the local spin energy and $J$ is the inter-qubit
coupling strength. We will say that the system displays SR--like
behaviour when we can identify some suitable figure of merit to
characterize the system response in the steady state, be it
dynamical or information--theoretic, such that it is nonmonotonic as
a function of the environmental noise strength. Dynamical figures of
merit for coupled, driven spin systems are typically magnetization
properties along a given direction \cite{HP07}, which provide a
suitable generalization of the standard measurements proposed for
single spin systems \cite{bu1,nancy}. Here we will focus on an
information theoretic approach and adopt the quantum mutual information,
which measures the total amount of correlations across any bipartition in the system \cite{Henderson}, as
the suitable figure of merit to quantify the presence of SR
\cite{igor}. Of particular interest for us is discerning whether there is steady state entanglement in the system. In the case of bipartite entanglement
this will be quantified by a suitable entanglement measure like, for
instance, the entanglement of formation \cite{eof} (see Sect. 2 for precise definitions).

We have organized the presentation of our results as follows. We
start by discussing the case of system qubits subject to pure
decay. We first revise the case of longitudinally coupled
qubits discussed in \cite{HP07} and revisit in some detail the basic
mechanisms yielding an inseparable mixed steady state. We show
that similar results hold when qubits couple via an exchange
interaction, which has been the object of recent interest in the
related context of noise assisted excitonic transport
\cite{aspuru,bio,olaya,fleming}. The next section deals with systems subject to
pure dephasing and we show that no steady-state correlations
will be displayed in this case. Finally we analyze the case
when the array is subject to both types of noise and re-evaluate the
emergence of a threshold, above which quantum correlations are non zero, 
in this situation, as well as the behaviour of the total correlation
content. The final section summarizes our findings and elaborates on
future work.

\section{Steady state entanglement in qubit chains subject to longitudinal decoherence (pure decay)}
The system to study is a chain of $N$ coupled qubits, each of them
interacting with independent, local,
\begin{figure}[ht]
{\centering
    \resizebox{0.75\columnwidth}{!}{ \includegraphics{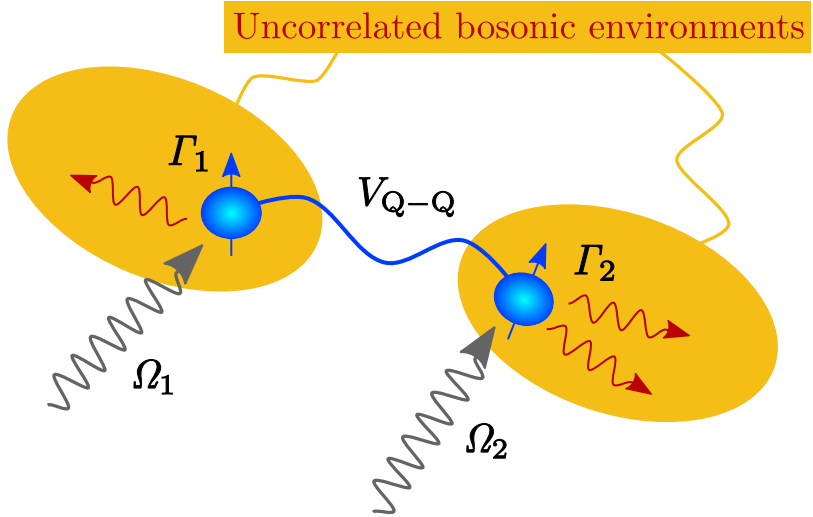} }
    \caption{Illustration of the generic set up. The simplest scenario if provided
		by an array of $N=2$ qubit systems, whose quantum state we denote by $\rho_{12}$, 
        with coherent interaction Hamiltonian $V_\mathrm{Q-Q}$ subject locally to an external driving of strength $\Omega_j$ and a decohering environment to which it couples with strength $\Gamma_j$.
		Note that the noise is local and therefore does not act as direct mediator
		of qubit interactions. \label{fig1}}
}
\end{figure}
harmonic baths and driven by an external field, so the total
system+environment Hamiltonian will be given by
\begin{equation}
H= -\frac{1}{2}\sum_{j=1}^N\omega^j_0\sigma_z^{j}+\sum_{j,k}\omega_k^j(a_k^j)^\dagger a_k^j+\sum_{j=1}^N\Omega_j(\sigma_+^je^{-i\omega_L^jt}+\mathrm{h.c.})+V_{\mathrm{Q-Q}}+V_{\mathrm{Q-Bath}},
\end{equation}
here $V_{\mathrm{Q-Q}}$ denotes the interqubit coupling Hamiltonian
and $V_{\mathrm{Q-Bath}}=\sum_{j=1}^N \hat{\xi}^j \otimes
X^j=\sum_{j,k}C_k \hat{\xi}^j\otimes[a_k^j+(a_k^j)^\dagger]$ is the
Hamiltonian term describing the interaction with the baths. Let us
consider the case first analyzed in \cite{HP07} where qubits exhibit
longitudinal coupling of the form
$V_{\mathrm{Q-Q}}=-J\sum_{j=1}^{N-1}\sigma_z^j\otimes\sigma_z^{j+1}$
and qubit--bath coupling $\hat{\xi}^j=\sigma_x^j$. We
adopt here the notation \cite{principles} when referring to transverse or longitudinal
coupling and note that a longitudinal decoherence allows for energy exchange
while a transverse coupling to the bath results solely in pure
dephasing.

Moving to a frame rotating with the external driving (by the
unitary transformation $U=\prod_{j=1}^Ne^{i\omega_L^j\sigma_z^j}$), we
arrive at the following master equation for the reduced density
matrix of the qubits
\begin{equation}\label{master1}
\frac{d\rho}{dt} =
-i\left(H_{\mathrm{eff}}\rho-\rho H_{\mathrm{eff}}^\dagger\right)
+ 2\sum_{j=1}^N\Gamma_j(\bar{n}_j+1)\sigma_-^j\rho\sigma_+^j
+ 2\sum_{i=1}^N\Gamma_j\bar{n}_j    \sigma_+^j\rho\sigma_-^j ,
\end{equation}
where we have introduced an effective, non-Hermitian term
\begin{equation}\label{Heff}
H_{\mathrm{eff}} = H_{\mathrm{coh}} - i\sum_{j=1}^N\Gamma_j(\bar{n}_j+1)\sigma_+^j\sigma_-^j
- i\sum_{j=1}^N\Gamma_j\bar{n}_j\sigma_-^j\sigma_+^j,
\end{equation}
with the coherent part of the evolution contained in
\begin{equation}
H_{\mathrm{coh}} = - \frac{1}{2}\sum_j\delta^j\sigma_z^{j}  +
\sum_{j=1}^N\Omega_j\sigma_x^j -
J\sum_{j=1}^{N-1}\sigma_z^j\otimes\sigma_z^{j+1} ,
\end{equation}
where $\delta^j$ is the detuning from the qubit transition
frequency, $\Omega_j$ denotes the Rabi frequency of the external
driving and $J$ is the strength of the coherent interqubit coupling.
The number of quanta in the local baths is given according to the
Bose-Einstein distribution
$\bar{n}_j = [\exp(\omega_0^j/k_\mathrm{B}T) - 1]^{-1}$.
In addition the decay
rates are expressed as $\Gamma_j=\pi \mathcal{J}(\omega_0^j)$ where
$\mathcal{J}(\omega)=\sum_kC_k^2\delta(\omega-\omega_k^j)$ is the spectral
density of each bath. This master equation
treatment is valid in the parameter regime $\Omega_j/\omega\ll1$, $\Gamma_j\bar{n}/\omega\ll1$, $\delta_j/\omega\ll1$ and $J/\omega\ll1$, where $\omega=\min\{\omega_0^j,\omega_c\}$  for a suitable bath's
frequency cut off $\omega_c$ \cite{cohen}. Note that in our study, decoherence rates will be
considered as ad-hoc parameters whose specific value is set by the
details of the qubit--bath interaction and the noise's spectral
properties so that these parameters provides with an effective
measure of the {\em noise strength} acting on the system. Moreover,
we will operate at $T=0$ given that the presence of a (realistic)
finite temperature does not modify the qualitative features we
discuss, but simply reduces the amplitude of the entanglement or
mutual information maximal values, as shown in \cite{HP07}.

\begin{figure}[ht]
{\centering
    \resizebox{0.75\columnwidth}{!}{ \includegraphics{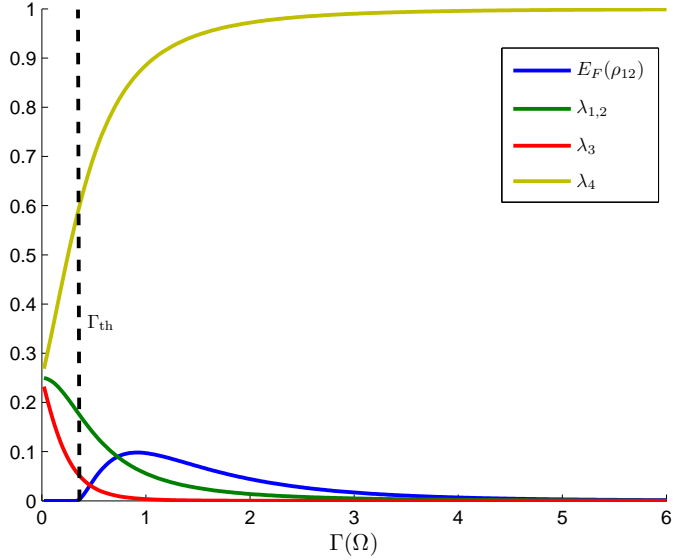} }
    \caption{Eigenvalues and entanglement content of the steady state given by equation (\ref{ss1}) as a function of the noise strength (for $J/\Omega=1.5$).  The dashed vertical line divides entangled and separable regimes. See main text for explanations.
    \label{fig2}}
}
\end{figure}

Let us consider the simplest case where the arrays consist of only
two qubits whose state we will denote by $\rho_{12}$, 
as depicted in figure 1, and let us assume, for simplicity, that 
$\Omega_1=\Omega_2=\Omega$ and $\Gamma_1=\Gamma_2=\Gamma$. At perfect tuning $\delta_j=0$
and zero temperature, the steady state of the master equation given by equation (\ref{master1})
can be computed analytically to be \cite{HP07}
\begin{equation}\label{ss1}
\rho^\mathrm{ss}_{12}=\frac{1}{k}\left(\begin{array}{cccc}
 t^2+4s^2r^2 ,& 2sr^2+irt ,& 2sr^2+irt ,& 2irs-r^2 \\
2sr^2-irt ,& t ,& r^2 ,& i r \\
2sr^2-irt ,& r^2 ,& t ,& i r \\
-2irs-r^2 ,& -i r ,& -i r ,& 1
\end{array}\right) \label{rho}
\end{equation}
where $k=3+2r^2+t^2+4r^2s^2$, $r=\Gamma/\Omega$, $s=J/\Omega$ and
$t=r^2+1$. This state is separable if, and only if, its partial
transpose is a positive operator \cite{horo}.  We find that
$\rho^\mathrm{ss}_{12}$ is entangled for values of the noise strength $\Gamma$ such that
\begin{equation}
    \Gamma>\Gamma_\mathrm{th},\quad \mathrm{where}\ \Gamma_\mathrm{th}=\frac{\Omega^2}{2J}.
    \label{thr}
\end{equation}
As a result, the systems exhibit quantum correlations in the steady
state only if the noise strength, encapsulated in the parameter
$\Gamma$, is above the threshold value $\Gamma_\mathrm{th}$.  Otherwise, the steady state
is fully separable. In order to shed light onto this rather
counterintuitive behaviour, it is useful to analyze the structural
form of the steady state density matrix by calculating its spectral
resolution. One can easily evaluate the eigenvalues of the density matrix
specifying the steady state equation (\ref{rho}) to be
\begin{eqnarray}
\lambda_{1,2}&=&\frac{1}{4 r^2 s^2+(1+t)^2} , \\
\lambda_{3,4}&=&\frac{1+r^2 \left(2+4 s^2\right)+
t^2\mp\sqrt{4 r^2 \left(1+s^2\right)+(t-1)^2} \sqrt{4 r^2 s^2+(1+t)^2}}{2 \left(4 r^2 s^2+(1+t)^2\right)}.
\end{eqnarray}
The corresponding expressions for the eigenvectors are not so
compact. Figure \ref{fig2} shows these eigenvalues as a function of
the noise strength as well as the steady state
entanglement as measured by the entanglement of formation
$E_F(\rho_{12})$. This quantity represents the minimal
possible average entanglement over all pure state decomposition
of $\rho$ and as such, its evaluation requires solving
a variational problem. However, in the case of two
qubits, the entanglement of formation has a simple closed 
form in terms of the so-called two-qubit concurrence, defined 
as $C(\rho)=\max\{0,\mu_1-\mu_2-\mu_3-\mu_4\}$, where the
$\mu_i$ are, in decreasing order, the eigenvalues of the matrix
$\rho\sigma_y\otimes\sigma_y\rho^\ast\sigma_y\otimes\sigma_y$, where $\rho^\ast$ 
is the matrix obtained by element complex conjugation of $\rho$ \cite{eof}. For any bipartite
qubit state,
\begin{equation}
E_F(\rho)=s\left(\frac{1+\sqrt{1-C^2(\rho)}}{2}\right),
\end{equation}
where $s(x) = -x \log_2x-(1-x)\log_2(1-x)$. We see in figure
2 that the entanglement of formation is zero for noise
values below threshold, while displaying the typical SR-like
profile for $\Gamma>\Gamma_\mathrm{th}$. When looking at the behaviour of
the spectral components, we observe that as $\Gamma$ increases, the
weights of some spectral components decrease and the chain tends to
localize in a certain eigenstate, in this case the one corresponding to
$\lambda_4$, which in the limit $\Gamma\rightarrow\infty$ is
actually the product state of the local Hamiltonian ground states.
That means the qubits tend to be in their individual ground states
as the effective decay rate becomes very large, as would be expected
intuitively. Interestingly, the approach to this separable
steady-state, and therefore the system's purity, is monotonic in the
noise strength $\Gamma$, despite the steady-state entanglement
exhibiting nonmonotonic SR--like behaviour.
This picture is in agreement with the behaviour already described in
\cite{plenio02} where, in the absence of dissipative noise in the
form of an additional heat bath, a thermally driven composite system
would approach a thermal steady state, while the presence of a
second heat bath at different temperature has the potential to make
the steady state mixture entangled. This behaviour may in fact be
quite general as closely related scenarios have now been identified
where steady state entanglement in facilitated by the presence of a
noisy channel \cite{hans,brandes,aguado}.

\section{XXYY Heisenberg Interaction}
It is worth exploring whether these results remain valid for other
kinds of qubit--qubit interactions. The most general coupling
between qubits is given by the Heisenberg Hamiltonian
\[
V_\mathrm{Hei}=-J_x\sigma_x\otimes\sigma_x-J_y\sigma_y\otimes\sigma_y-J_z\sigma_z\otimes\sigma_z.
\]
However, the evolution equation for this general Hamiltonian can no
longer be written in terms of a time independent effective Hamiltonian, because the
transformation $U=\prod_{j=1}^Ne^{i\omega_L^j\sigma_z^j}$ does not
commute with $V_{XXYY}$. As long as the frequency of the driving is
the same for each qubit $\omega_L^j=\omega_L$, the most general
Heisenberg-type of Hamiltonian which is time--independent in the
rotating picture is the so-called XXYY interaction:
\[
V_\mathrm{XXYY} =
- J_\perp(\sigma_x\otimes\sigma_x+\sigma_y\otimes\sigma_y)
- J_\parallel\sigma_z\otimes\sigma_z .
\]
It is easy to check that $U V_\mathrm{XXYY}U^\dagger=V_\mathrm{XXYY}$.

\begin{figure}[!ht]
{\centering
    \resizebox{0.75\columnwidth}{!}{ \includegraphics{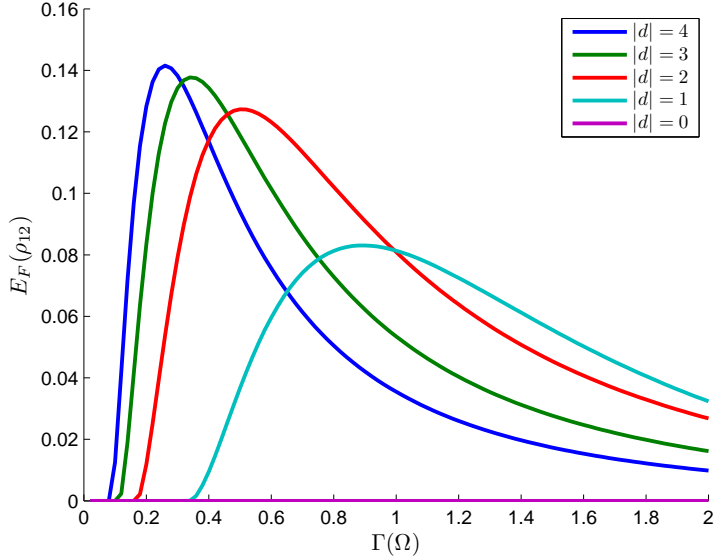} }
    \caption{Increasing behaviour of the steady-state entanglement for
    $|d|$ growing from $0$ to $4$. The system is separable when
    $d=0$ (isotropic Heisenberg model). When $d$ increases, and the
    term $XXYY$ dominates, the range of values of $\Gamma$ for which
    the steady state is separable shrinks, while the maximum
    achievable steady state entanglement grows.
    \label{fig3}}
}
\end{figure}

Again for two qubits with zero detuning at zero temperature, the steady state can be
found analytically:
\[
\rho^\mathrm{ss}_{12}=\left(\begin{array}{cccc}
 t^2+4d^2r^2 ,& 2dr^2+irt ,& 2dr^2+irt ,& 2ird-r^2 \\
2dr^2-irt ,& t ,& r^2 ,& i r \\
2dr^2-irt ,& r^2 ,& t ,& i r \\
-2ird-r^2 ,& -i r ,& -i r ,& 1
\end{array}\right) ,
\]
where $d=(s_\perp-s_\parallel) = (J_\perp-J_\parallel)/\Omega$. This
is the same as the steady state in (\ref{ss1}), with $s$ replaced by
the parameter $d$. As a result, we obtain an entangled steady-state
for $\Gamma_{\mathrm{th}}<\Gamma$, where the threshold this
time is
\[
\Gamma_{\mathrm{th}}=\frac{\Omega^2}{2|d|} .
\]
This interesting formula shows that for the isotropic Heisenberg
model $J_\parallel=J_\perp$, the steady state remains separable.
Furthermore, increasing $|d|$ (the dominance of XXYY over ZZ or
conversely in an anisotropic Heisenberg model) increases the maximum
steady-state entanglement, as illustrated in figure \ref{fig3}).
This is sensible as the interaction becomes more entangling the
further it deviates from a product of local Hamiltonians.

\section{Steady state entanglement under transverse decoherence (pure dephasing)}
Apart from processes describing emission or absorption of quanta,
qubit implementations can undergo energy conserving, purely dephasing
processes, where only the coherences are affected and the populations
remain unchanged. This is the situation encountered when the
interaction term with the bath is mediated by an operator that
commutes with the qubit Hamiltonian.  Given the system Hamiltonian
of one qubit $H=(\omega_0/2)\sigma_z$, a feasible qubit--bath
interaction is $V=\sigma_zX$ where the force operator is
$X=\sum_kC_k(a_k+a_k^\dagger)$. This leads, under the customary
approximations, to the master equation
\begin{equation}
\frac{d\rho}{dt} = -i[H,\rho] + 2\gamma(\sigma_z\rho\sigma_z - \rho) ,
\end{equation}
where
\[
\gamma = \frac{1}{2}\lim_{\omega\rightarrow0}S_X(\omega)
 = \pi\lim_{\omega\rightarrow0}\mathcal{J}(\omega)\coth\left(\frac{\omega}{2k_\mathrm{B}T}\right)
\]
for a thermal bath with spectral density $\mathcal{J}(\omega)=\sum_kC_k^2\delta(\omega-\omega_k)$.

This process can be embedded in our framework provided that the Rabi
frequencies of the external driving $\Omega_j$ and the interqubit
coupling $J$ are small enough in comparison with the free evolution
frequencies \cite{HP07}. The master equation in the rotating picture
for $N$ qubits is then given by
\begin{equation}
\frac{d\rho}{dt} = -i[H_{\mathrm{coh}},\rho] +
2\sum_{j=1}^N\gamma_j(\sigma_z^j\rho\sigma_z^j-\rho) \label{deco}.
\end{equation}
In this situation the system no longer exhibits quantum
correlations in the steady state. Indeed, the steady state of
equation (\ref{deco}) is unique and it is the completely mixed state
$\rho^{\mathrm{ss}}=\mathds{1}/N$. This result follows from theorem
5.2 in \cite{Sp80-Fr78} which asserts that given a steady state
$\rho^{\mathrm{ss}}$ of a master equation written in the standard
form
\[
\frac{d\rho}{dt}=-i[H,\rho]+\sum_j\left(V_j\rho V_j^\dagger-\frac{1}{2}V_j^\dagger V_j\rho-\frac{1}{2}\rho V_j^\dagger V_j\right) ,
\]
such that $\mathrm{rank}(\rho^{\mathrm{ss}})=N$, the given steady state is
unique if the only operators commuting with $H$ and every $V_j$ are
multiples of the identity.  This is true for our case since the only operators
that commute with both $\sigma_x$ and $\sigma_z$ (and therefore with
$\sigma_y$ as well because of the Jacobi identity) are proportional to the
identity because of Schur's lemma.

\section{SR phenomena under both longitudinal and transverse decoherence}
The situation changes if we have both emission/absorption and pure dephasing processes
in our system, as is the case in many solid state qubit implementations
(see for example \cite{schoen} and references therein).
Then the master equation is modified according to
\begin{equation}
\frac{d\rho}{dt} = -i\left(H_{\mathrm{eff}}\rho-\rho H_{\mathrm{eff}}^\dagger\right) + \sum_j
2\Gamma_j(\bar{n}+1)\sigma_-^j\rho\sigma_+^j + \sum_j
2\Gamma_j\bar{n}\sigma_+^j\rho\sigma_-^j + \sum_j 2 \gamma_j
\sigma_z^j\rho\sigma_z^j ,
\end{equation}
where $\gamma_j$ denotes the pure dephasing rates, and
$H_{\mathrm{eff}}$ is given by the same equation (\ref{Heff}) with an
additional term of the form $-i \sum_j \gamma_j$ accounting for
pure dephasing.

\begin{figure}[ht]
{\centering
    \resizebox{0.75\columnwidth}{!}{ \includegraphics{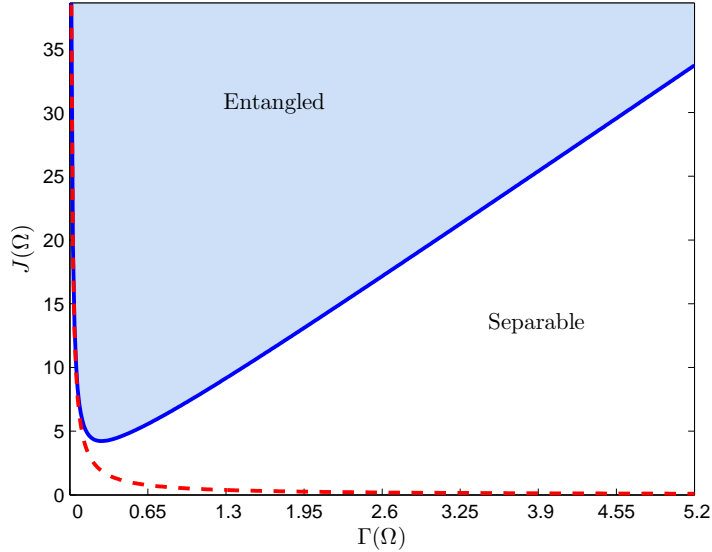} }
    \caption{Steady state entangling behaviour: $J$ vs. $\Gamma$ without pure dephasing
    (red dashed line), and with pure dephasing (blue solid line).
    The blue shading shows the entangled zone in the presence of pure dephasing.
    \label{fig4}}
}
\end{figure}

\begin{figure}[!ht]
{\centering
    \resizebox{0.75\columnwidth}{!}{\includegraphics{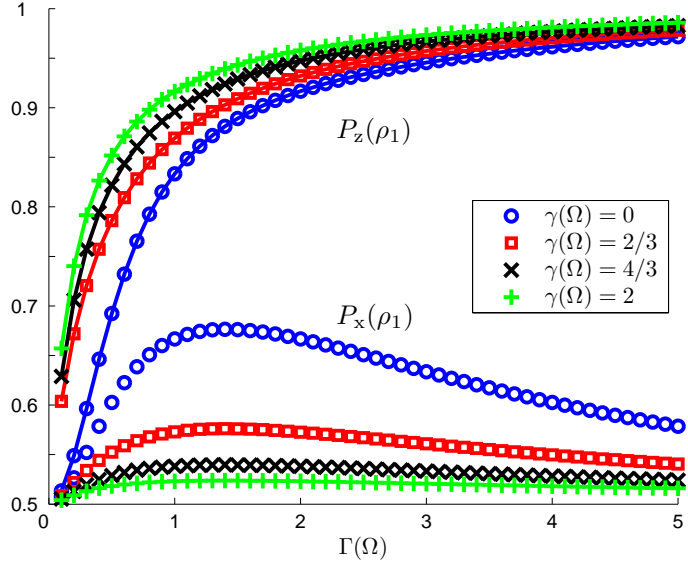} }
    \caption{Illustration of the combined action of transverse and longitudinal
    decoherence on an array of $N=2$ qubits with a $ZZ$ interqubit Hamiltonian.
		Increasing the pure dephasing rate keeps the {\em delocalization}
		probability of single qubits $P_\mathrm{x}$ close to $1/2$ so that the qubits
		are in a state close to an eigenstate of their local Hamiltonians ($\sigma_z$)
		which reduces the entangling power of the coherent $ZZ$ coupling.
    \label{fig5}}
}
\end{figure}

Consider $T=0$ (also assuming that each $\gamma_j$ is finite).
For two qubits in the absence of pure dephasing, the threshold for an
entangled steady state was given in (\ref{thr}) to be \cite{HP07}
\[
\rho_{12}^\mathrm{ss}\ \mathrm{entangled}\Leftrightarrow \frac{1}{2s} < r.
\]
We now want to do the same analysis with the pure dephasing term added.
For simplicity let us assume $\gamma=\Gamma$.  The result is now more
complicated as shown in figure \ref{fig4} where $J$ is plotted against $\Gamma$
($s$ vs. $r$) without pure dephasing (red dashed line), and with pure dephasing
(blue solid line dividing entangled and separable zones).
For illustrative purposes we have included values of $J$ substantially
larger than $\Omega$, whilst still adhering to the weak-coupling assumptions of
the master equation.
There are several differences between the two cases.
For both cases, the steady state is separable for very small $\Gamma$,
so it is necessary to include some appreciable amount of noise to produce
entanglement (SR--like behaviour).
However, in the presence of pure dephasing, only a finite amount of additional
noise is required to come back to the separable regime.  (Recall that an infinite
amount of noise is required in the absence of pure dephasing.)
Furthermore, in the presence of pure dephasing the SR--like behaviour
disappears completely for small $J$, where the steady state remains separable.
At odds with the monotonic result (\ref{thr}) of \cite{HP07} (red dashed
line of figure \ref{fig4}), the nonmonotonicity of the blue solid threshold in
figure \ref{fig4} (note the minimum point) reflects in some sense the
competition between both processes. Consider the simplest case where qubits
were subject to a $ZZ$ coupling. In figure \ref{fig5}, where we plot the
behaviour of the probabilities $P_{z}$ (localization) and $P_{x}$
(delocalization) for each qubit to be in an eigenstate of the
corresponding Pauli operator (i.e., $\sigma_z$ or $\sigma_x$).  The closer the local states are to an
eigenstate of $\sigma_z$, the less effective the coherent $ZZ$ coupling
is and the global state tends to be separable.  The localization
probability increases steadily with the transverse decay rate;
however, the delocalization probability is maximal for some optimal
noise strength provided that pure dephasing is not too large. The region
around the maximum delocalization probability coincides with
the maximum steady state entanglement.

The exact form of the threshold is complicated \cite{note}. An
approximate solution for it is
$s\simeq\frac{1}{2r}+\frac{32r+2}{5}$. So that the steady state is
entangled for a noise strength in, approximately, the range
\[
-1/32+(5/64)s-(1/64)\sqrt{25s^2-20s-316} < r < -1/32+(5/64)s+(1/64)\sqrt{25s^2-20s-316}
\]
Note that the lower bound is always higher
than the previous one, $1/2s$.

\begin{figure}[ht]
{\centering
    \resizebox{0.75\columnwidth}{!}{ \includegraphics{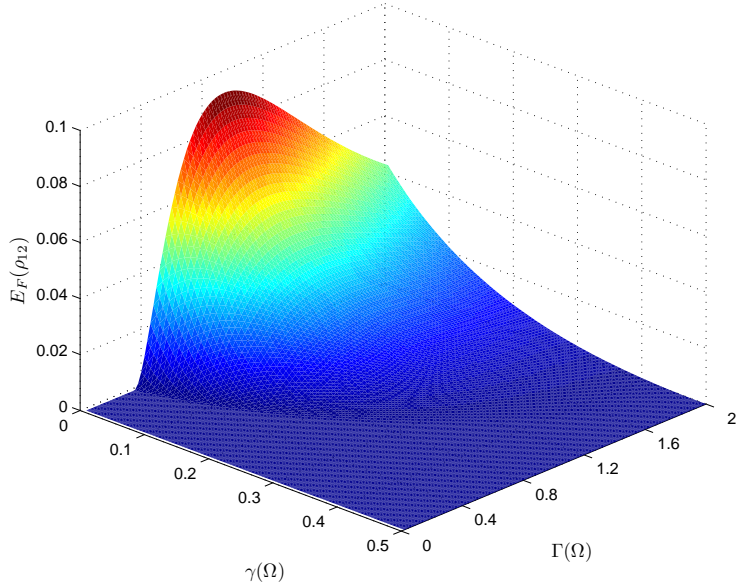} }
    \caption{Quantum correlations in a two-qubit array as quantified by the entanglement of formation as a function of
    the noise parameters $\Gamma$ and $\gamma$ for $J/\Omega=1.5$.
    Note the monotonous negative effect when $\gamma$ increases so
    that as the pure dephasing rate increases, the larger it becomes
    the threshold value for the transverse decoherence until reaching
    a critical value $\gamma_{c}$ for which no entanglement survives
    in the steady state.
    \label{fig6}}
}
\end{figure}

\begin{figure}[!ht]
{\centering
    \resizebox{0.75\columnwidth}{!}{\includegraphics{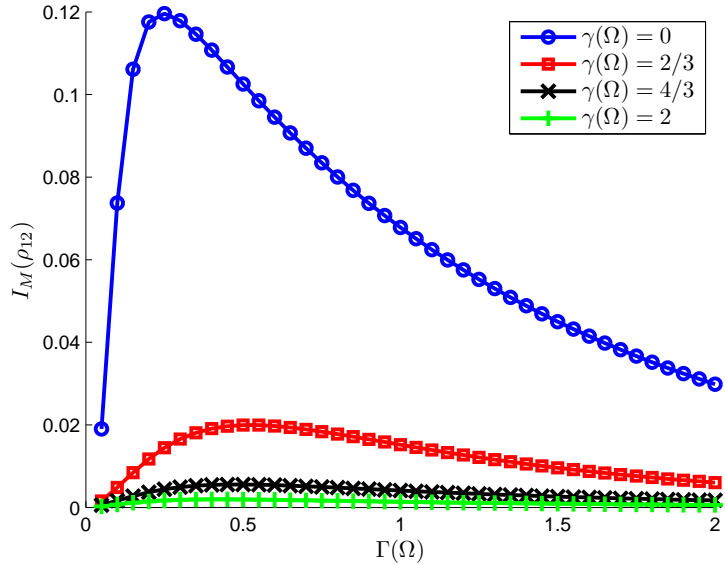} }
    \caption{Quantum mutual information $I_M(\rho_{12})$ between qubits 1 and 2 in an $N=6$ array
    for different values of the dephasing rate $\gamma$ and variable
    transverse decoherence rate $\Gamma$ for $J/\Omega=1.5$. The same qualitative behaviour
    is observed for other pairs of neighbouring qubits or across
    different bipartitions within the system.  The quantum mutual information is
    maximized for an optimal decay rate $\Gamma$ above which the total
    correlations in the system start to degrade. The larger the
    dephasing rate acting locally on the qubit systems, the smaller
    the value of the maximum correlation content becomes.
    \label{fig7}}
}
\end{figure}

For the case of arbitrary values of $\gamma$ and $\Gamma$, the
steady-state correlations are diminished as $\gamma$ increases. This
can be visualized in the behaviour of bipartite entanglement, as illustrated in figure \ref{fig6}.
In order to construct a proper information theoretic measure of SR,
we consider the quantum mutual information $I_M$, defined
for a general bipartite system AB as $I_M(\rho_{AB}) =
S(\rho_A)+S(\rho_B)-S(\rho_{AB})$, where the states $\rho_{A,B}$ are the local
states $\rho_{A,B} = \mathrm{Tr}_{B,A}(\rho_{AB})$. This function quantifies the
total correlation content in any bipartite system \cite{Henderson} and
is depicted in figure \ref{fig7} for the case of a chain of 6 qubits.
There we evaluate the mutual information of the first two
qubits in the chain but should stress that the same qualitative
behaviour, where total correlations are maximized for some optimal, intermediate value of the decay rate,
are obtained when evaluating the quantum mutual information
across any other bipartition in the chain. We note that having a nonzero 
detuning $\delta\ne0$ does not change the
shape of the entanglement function, but its magnitude is reduced;
the effect being small for small detuning, say $\delta^j \sim
10^{-3}-10^{-2}$.


\section{Conclusion}
We have discussed the emergence of SR--like effects in composite
spin systems subject to both decay (longitudinal decoherence) and pure dephasing
(transverse decoherence).
We have shown that quantum correlations vanish in the steady state if the local
noise is purely transverse. Under pure dephasing, the system
evolves into a maximally mixed state. When transverse and
longitudinal noise are simultaneously present, for fixed dephasing,
one encounters a decay rate threshold for steady--state entanglement to
exist, provided that dephasing noise remains moderate.
In the special case where $\Gamma=\gamma$, an approximate analytical
threshold condition can be derived.

To quantitatively characterize the presence of stochastic resonance,
we compute the total correlation content of the system, as
quantified by its mutual information for bipartitions of arbitrary
size, and show that it is a non-monotonic function of the decay rate
$\Gamma$ (for fixed $\gamma$).  We therefore argue that the system
displays SR as measured by an information--theoretic figure
of merit.

In the present work we have focussed our interest on the
correlations content and, in particular, on the entanglement content
of the steady state. These results show that a noisy environment
does not just monotonically degrade the amount of entanglement but
can act, for a suitable noise level, as a {\em purifying} mechanism
that prevents the system from thermalizing --- unless the noise
becomes too strong \cite{plenio02}.

It seems to become more and more clear that the presence of
environmental noise, far from being always detrimental, may actually
be instrumental in the optimization of certain processes. So far
this is more clear--cut in the case of processing classical
information, as in the original SR concept of amplifying a weak
signal, or the quantum setting in the context of the assisting
transport of excitons across spin networks. Those can model for
instance natural phenomena such as exciton transport in light--harvesting
complexes \cite{aspuru,bio,olaya,fleming}.  However, there
are already indications that noise may also assist the transfer of
quantum information, as exemplified by the enhanced fidelity in the
transmission of quantum states demonstrated in \cite{difranco}. It
would be extremely important to generalize this result and to
clearly identify the conditions under which not only quantum
correlations, but also the transfer of quantum information, can be
effectively assisted by noise.

\begin{acknowledgement}
    This work was supported by the EU STREP
    project CORNER and the EU Integrated project on \emph{Qubit
    Applications} QAP. AR acknowledges support from a University of
    Hertfordshire Fellowship. We are grateful to Martin Plenio and Shash Virmani for
    numerous and inspiring discussion on the topic of this paper and to
    Igor Goychuk for his comments during the SR2008 meeting in Perugia.
\end{acknowledgement}


\begin{thebibliography}{}
\bibitem{note} The exact form of the threshold is provided by the only real root of the polynomial:
   \begin{eqnarray*}
   &\mbox{}& 20480 r^4 s^{12} + 4096 (r^2 (-1 + 20 r^2 + 205 r^4)) s^{10} +
256 (-1 - 280 r^2 - 7208 r^4 - 64560 r^6 + 17470 r^8) s^8\\
   &-& 1792 (2 + 295 r^2 + 10634 r^4 + 149910 r^6 + 764420 r^8 + 964565 r^{10}) s^6\\
   &-& 784 (24 + 2600 r^2 + 97542 r^4 + 1665144 r^6 + 13511368 r^8 +
   48202000 r^{10} + 59489675 r^{12}) s^4\\
   &-& 5488 (2 + 35 r^2)^2 (2 + 115 r^2 + 2122 r^4 + 15671 r^6 +
   50168 r^8 + 58352 r^{10}) - 2401 (2 + 43 r^2 + 140 r^4)^4=0,
   \end{eqnarray*} for which it does not exist a general
closed solution.
\end{thebibliography}
\end{document}